\documentclass[10pt,a4paper,twocolumn,english,pra,showpacs,noshowkeys,superscriptaddress]{revtex4}

\usepackage[LGR,T1]{fontenc}
\usepackage[latin1]{inputenc}
\usepackage{color}
\usepackage{babel}
\usepackage{amsmath}
\usepackage{amssymb}
\usepackage{graphicx}
\usepackage{hyperref}
\usepackage{color}
\usepackage{babel}
\usepackage{amstext}
\usepackage{times}

\hypersetup{colorlinks,citecolor=blue,filecolor=black,linkcolor=red,
urlcolor=blue,pdftex} \makeatletter

\@ifundefined{textcolor}{} {
 \definecolor{BLACK}{gray}{0}
 \definecolor{WHITE}{gray}{1}
 \definecolor{RED}{rgb}{1,0,0}
 \definecolor{GREEN}{rgb}{0,1,0}
 \definecolor{BLUE}{rgb}{0,0,1}
 \definecolor{CYAN}{cmyk}{1,0,0,0}
 \definecolor{MAGENTA}{cmyk}{0,1,0,0}
 \definecolor{YELLOW}{cmyk}{0,0,1,0}
}
\makeatother

\begin{document}

\title{Localized modes in quasi-2D Bose-Einstein condensates with spin-orbit
and Rabi couplings}
\author{Luca Salasnich}
\affiliation{Dipartimento di Fisica e Astronomia ``Galileo Galilei'' and CNISM, Universit\`a
di Padova, Via Marzolo 8, 35131 Padova, Italy}
\affiliation{Istituto Nazionale di Ottica (INO) del Consiglio Nazionale delle Ricerche
(CNR), Sezione di Sesto Fiorentino, Via Nello Carrara, 1 - 50019 Sesto
Fiorentino, Italy}
\author{Wesley B. Cardoso}
\affiliation{Instituto de F\'isica, Universidade Federal de Goi\'as, 74.001-970,
Goi\^ania, Goi\'as, Brazil}
\author{Boris A. Malomed}
\affiliation{Department of Interdisciplinary Studies, School of Electrical Engineering,
Faculty of Engineering, Tel Aviv University, Tel Aviv 69978, Israel}

\begin{abstract}
We consider a two-component pancake-shaped, i.e., effectively
two-dimensional (2D), Bose-Einstein condensate (BEC) coupled by the
spin-orbit (SO) and Rabi terms. The SO coupling adopted here is of the mixed
Rashba-Dresselhaus type. For this configuration, we derive a system of two
2D nonpolynomial Schrödinger equations (NPSEs), for both attractive and
repulsive interatomic interactions. In the low- and high-density limits, the
system amounts to previously known models, namely, the usual 2D
Gross-Pitaevskii equation, or the Schrödinger equation with the nonlinearity
of power $7/3$. We present simple approximate localized solutions, obtained
by treating the SO and Rabi terms as perturbations. Localized solutions of
the full NPSE system are obtained in a numerical form. Remarkably, in the
case of the attractive nonlinearity acting in free space (i.e., without any
2D trapping potential), we find parameter regions where the SO and Rabi
couplings make 2D fundamental solitons dynamically stable.
\end{abstract}

\maketitle

\section{Introduction}

The Bose-Einstein condensates (BECs) have become an important ground for the
study of macroscopic quantum phenomena. The first experimental realizations
in 1995 were implemented using rubidium \cite{AndersonSC95}, sodium \cite%
{DavisPRL95}, and lithium \cite{BradleyPRL95} atoms, which motivated a great
deal of work \cite{DalfovoRMP99} on topics such as the Anderson localization
of matter waves \cite{BillyNAT08,RoatiNAT08}, production of bright \cite%
{KhaykovichSCi02,StreckerNAT02,CornishPRL06, MarchantNC13} and dark solitons
\cite{BurgerPRL99}, dark-bright complexes \cite{BeckerNP08}, vortices \cite%
{MatthewsPRL99} and vortex-antivortex dipoles \cite%
{NeelyPRL10,FreilichSC10,SemanPRA10,MiddelkampPRA11}, persistent flows in
the toroidal geometry \cite{RyuPRL07,RamanathanPRL11}, skyrmions \cite%
{WusterPRA05}, emulation of gauge fields \cite{LinNP11} and spin-orbit (SO)
coupling \cite{LinNAT11}, quantum Newton's cradles \cite{KinoshitaNAT06},
\textit{etc}.

The recent realization of the artificial SO coupling in a neutral atomic BEC
\cite{LinNAT11} was an incentive for many more works, dealing, in
particular, with vortex structures in rotating SO-coupled BECs \cite%
{HoPRL11,XuPRL11,RadicPRA11,ZhouPRA11} and trapped 2D atomic BEC with
spin-independent interactions in the presence of the isotropic SO coupling
\cite{SinhaPRL11,HuPRL12}, which shows that, for weak interactions, two
types of half-vortex solutions with different winding numbers occur.
Further, in the weakly interacting regime realized for the two-component
Bose gas in a 2D harmonic-oscillator (HO)\ trap, subject to isotropic SO
coupling of the Rashba type, it was found that the condensate's ground state
has a half-quantum angular momentum vortex configuration with spatial
rotational symmetry and skyrmion-type spin texture \cite{RamachandhranPRA12}%
. The introduction of order by disorder in Rashba-SO-coupled BECs was
demonstrated in Ref. \cite{BarnettPRA12}. In Ref. \cite{DengPRL12}, an
experimental scheme was proposed for the creation of the SO coupling in
spin-3 Cr atoms, using the Raman illumination, and the ground-state
properties of that model were studied. In Ref. \cite{WenPRA12}, the
ground-state properties of a weakly trapped spin-1 BEC with the SO coupling
were studied by means of numerical and analytical methods in an external
Zeeman field. The existence of complex, antiferromagnetically ordered
(striped), ground states in the 1D SO-coupled system with the repulsive
interactions and external HO trap was demonstrated in Ref. \cite{Zezyu}.

Bright solitons in the BEC with the SO coupling were introduced in Refs.
\cite{XuPRA13,AchilleosPRL13,KartashovPRL13,SalasnichPRA13} and \cite%
{SakaguchiPRE14,LobanovPRL14}, in the 1D and 2D geometries, respectively. In
particular, in Ref. \cite{SalasnichPRA13} localized modes in dense repulsive
and attractive BECs with the spin-orbit and Rabi couplings were
investigated; Ref. \cite{KartashovPRL13} reported a diversity of stable gap
solitons in the SO-coupled BEC subject to the action of a spatially periodic
Zeeman field; employing a multiscale-expansion method, bright-solitons
families and three different dark soliton families were found, respectively
in Refs. \cite{AchilleosPRL13} and \cite{AchilleosEPL13} (the dark solitons
could be supported by either a constant or a spatially modulated background
density). As concerns 2D bright solitons, an unexpected results was reported
in Ref. \cite{SakaguchiPRE14}: two different families of vortex solitons,
namely semi-vortices (with topological charges $m=0$ and $\pm1$ in the two
components) and mixed modes (which combine $m=0$ and $\pm1$ in each
component) are \emph{stable} in the 2D binary BEC with the Rashba coupling
in the \emph{free space}, without the support of any trapping potential,
while it commonly believed that all 2D free-space solitons are unstable in
the presence of the cubic attractive nonlinearity due to the occurrence of
the critical collapse in the same setting \cite{Berge'}. Various families of
2D localized solutions, including multipole and half-vortex solitons
featuring various symmetries with respect to the parity and time reversal in
a lattice created by the Zeeman field were reported in Ref. \cite%
{LobanovPRL14} for the mixed Rashba-Dresselhaus SO coupling.

The stability of plane waves in the the two-dimensional SO-coupled BEC was
studied analytically in Refs. \cite{HeEJPD13,HePRA13}. Vortex-lattice
solutions to the coupled Gross-Pitaevskii (GP) equations with the SO
coupling and optical-lattice potential were reported in Ref. \cite%
{SakaguchiPRA13}. Vortex dynamics in the SO-coupled BEC was studied in Ref.
\cite{FetterPRA14}, while Ref. \cite{LiJLTP14} addresses the existence of
the vortex-antivortex-pair lattices in the BEC with the Rashba SO coupling.
Recently, Ref. \cite{ChengPRA14} reported, by means if numerical and
variational methods, the localization of a noninteracting and weakly
interacting SO-coupled BEC in a quasiperiodic bichromatic optical-lattice
potential, confirming the existence of stationary localized states in the
presence of the SO and Rabi couplings for equal numbers of atoms in the two
components. In Ref. \cite{JinJPB14}, transitions of ground states, induced
by the zero-momentum coupling in the Rashba-SO-coupled BEC with pseudospin
1/2 were investigated in the presence of the confining HO potential.
Fragmentation of SO-coupled spinor BEC was studied in Ref. \cite{SongPRA14},
and Anderson localization of cold atomic gases with the effective spin-orbit
interaction in a quasiperiodic optical lattice was theoretically
investigated in Ref. \cite{ZhouPRA13}.

Most works dealing with patterns produced by the SO coupling used GP
equations with the cubic nonlinearity. However, it is known that a
sufficiently strong transverse confinement leads to effective nonlinearity
which is different from cubic, thus given rise to nonpolynomial Schrödinger
equations (NPSEs) \cite{SalasnichLP02,Delgado}. In particular, in Ref. \cite%
{SalasnichPRA13} a system of coupled 1D NPSEs was derived from the full 3D
two-component GP equations including the SO and Rabi couplings. In the case
of the attractive nonlinearity, this effective 1D system features essential
reduction of the collapse threshold, under the action of the SO and Rabi
couplings, suggesting an expansion of the range of physical parameters where
experiments may reveal new self-trapped modes. The variety of the
above-mentioned theoretical results obtained for the SO-coupled BEC in the
effectively 2D geometry suggests the relevance of deriving an effective 2D
NPSE system, starting from the full set of 3D GP equations with the SO and
Rabi couplings, and taking into regard the strong confinement in the
direction perpendicular to the 2D ``pancake''. The derivation of such a
system, and analysis of localized solutions (solitons) predicted by it, is
the subject of the present work.

The paper is organized as follows. The derivation of the 2D NPSEs system is
presented in Sec. II. along with simple analytical approximations for
localized modes produced by these equations, and the low- and high-density
limit cases. Numerical results for the localized states, obtained in the
system with the self-attractive and self-repulsive interactions are reported
in Sec. III. The paper is concluded by Sec. IV.

\section{Nonpolynomial Schrödinger equations for the spin-orbit-coupled
condensate}

The single-particle Hamiltonian with the SO-coupling term (of the mixed
Rashba-Dresselhaus type), which can be implemented in BEC, is
\begin{equation}
\hat{h}_{\mathrm{SP}}=\left[\frac{\hat{\mathbf{p}}^{2}}{2m}+U(\mathbf{r}) %
\right]+\frac{\hbar\Omega}{2}\sigma_{x}-\frac{k_{L}}{m}\hat{p}_{x}\sigma_{z},
\label{SP}
\end{equation}
where $\hat{\mathbf{p}}=-i\hbar(\partial_{x},\partial_{y},\partial_{z})$ is
the momentum operator, $U(\mathbf{r})$ is a trapping potential, $k_{L}$ is
the recoil wavenumber induced by the interaction with the laser beams, $%
\Omega$ is the frequency of the Raman coupling, which is responsible for the
linear mixing between the two states, and $\sigma_{x,y,z}$ are the Pauli
matrices.

\subsection{Derivation of the model}

The dilute SO- and Rabi-coupled binary BEC, confined in the $z$ direction by
a tight HO potential with trapping frequency $\omega _{z}$, and in the $(x,y)
$ plane by a generic loose potential $V(x,y)$, is governed by the system of
3D GPEs for macroscopic wave functions $\Psi _{k}(x,y,z,t)$ of the two
atomic states ($k=1,2$):
\begin{eqnarray}
&&i\partial _{t}\Psi _{k}=\left[ -\frac{1}{2}\nabla ^{2}+V(x,y)+\frac{1}{2}%
z^{2}+(-1)^{k-1}i\gamma \partial _{x}\right.   \notag \\
&&+\left. \sqrt{2\pi }g_{k}|\Psi _{k}|^{2}+\sqrt{2\pi }g_{12}|\Psi
_{3-k}|^{2}\right] \Psi _{k}+\Gamma \Psi _{3-k},  \label{3DGP-SORC}
\end{eqnarray}%
where the lengths, time, and energy are measured in units of $a_{z}=\sqrt{%
\hbar /(m\omega _{z})}$, $\omega _{z}^{-1}$, and $\hbar \omega _{z}$,
respectively. Here $g_{k}\equiv \sqrt{2\pi }(2a_{k}/a_{z})$, $g_{12}\equiv
\sqrt{2\pi }(2a_{12}/a_{z})$ are strengths of the intra- and inter-species
interactions, where $a_{k}$ and $a_{12}$ are the respective scattering
lengths, while $\gamma =k_{L}a_{z}$ and $\Gamma =\Omega /(2\omega _{z})$ are
dimensionless strengths of the spin-orbit and Rabi couplings, respectively.
The time-dependent number of atoms in the $k$-th state is $N_{k}(t)=\int
\int \int dxdydz|\Psi _{k}(x,y,z,t)|^{2}$, the constant total number of
atoms being $N=N_{1}(t)+N_{2}(t)$.

In most cases, a reasonable assumption is that strengths of the nonlinear
interactions between different atomic states are equal, $g_{1}=g_{2}=g_{12}%
\equiv g$ \cite{Pitaevskii03}. Under this condition, we aim to construct
stationary states with chemical potential $\mu $, by setting
\begin{equation}
\Psi _{k}(x,y,z,t)=\text{$\psi $}_{k}(x,y,z)\ e^{-i\text{$\mu $}t}
\label{st1}
\end{equation}%
in Eq. (\ref{3DGP-SORC}). The resulting equations for stationary fields $%
\psi _{1,2}(x,y,z)$ are compatible with the restriction
\begin{equation}
\psi _{1}^{\ast }(x,y,z)=\psi _{2}(x,y,z),  \label{st2}
\end{equation}%
which leads to the single stationary equation:
\begin{eqnarray}
\mu \Phi  &=&\left[ -\frac{1}{2}\nabla ^{2}+V(x,y)+\frac{1}{2}z^{2}\right.
\notag \\
&+&\left. i\gamma \partial _{x}+\sqrt{2\pi }g|\Phi |^{2}\right] \Phi +\Gamma
\Phi ^{\ast },  \label{static_3D}
\end{eqnarray}%
where we set $\Phi (x,y,z)\equiv \sqrt{2/N}\psi _{1}(x,y,z)=\sqrt{2/N}\psi
_{2}^{\ast }(x,y,z)$, so that $\int \int \int dxdydz|\Phi (x,y,z)|^{2}=1$.

To simplify the stationary 3D problem, we adopt the usual factorized ansatz
for the wave functions which are strongly confined in the direction of $z$,
and weakly confined in the plane of ($x,y$):
\begin{equation}
\Psi _{k}(x,y,z,t)=\frac{\exp \left[ -(1/2)\eta _{k}^{2}(x,y,t)z^{2}\right]
}{\pi ^{1/4}\sqrt{\eta _{k}(x,y,t)}}F_{k}(x,y,t),  \label{ansatz}
\end{equation}%
where $\eta _{k}(x,y,t)$ and $F_{k}(x,y,t)$ are the axial width and planar
wave function, respectively, the latter normalized by conditions
\begin{equation}
\int \int dxdy|F_{k}(x,y)|^{2}=1.  \label{norma}
\end{equation}

Inserting ansatz (\ref{ansatz}) into the energy functional which produces
Eq. (\ref{static_3D}), performing the integration in the transverse plane,
and neglecting, as usual, derivatives of $\eta_{k}(x,y)$, we derive the
corresponding effective energy functional, which reduces 3D equations (\ref%
{3DGP-SORC}) to a system of two effectively 2D equations:
\begin{eqnarray}
i(F_{k})_{t} & = & \left[-\frac{1}{2}(\partial_{x}^{2}+\partial_{y}^{2})
+V(x,y)+(-1)^{k-1}i\gamma\partial_{x}\right.  \notag \\
& + & \left.\frac{1}{4}\left(\frac{1}{\eta_{k}^{2}}+\eta_{k}^{2}\right) +%
\frac{g}{\eta_{k}}|F_{k}|^{2}\right]F_{k}+\Gamma F_{3-k},  \label{full_2D}
\end{eqnarray}
\begin{equation}
\eta_{k}^{4}=1+g|F_{k}|^{2}\eta_{k}.
\end{equation}
Further, stationary 2D solutions may be looked for as
\begin{equation}
F_{k}(x,y,t)=e^{-i\mu t}f_{k}(x,y)  \label{Fk}
\end{equation}
\begin{equation}
f_{1}(x,y)=f_{2}^{\ast}(x,y)\equiv f(x,y),\;\eta_{1}(x,y)=\eta_{2}(x,y)
\equiv\eta(x,y)  \label{Fk2}
\end{equation}
{[}cf. Eqs. (\ref{st1}), (\ref{st2}){]}, with complex function $f(x,y)$ and
real one $\eta(x,y)$ obeying the following equations:

\begin{eqnarray}
\mu f & = & \left[-\frac{1}{2}(\partial_{x}^{2}+\partial_{y}^{2})+V(x,y)
+i\gamma\partial_{x}\right.  \notag \\
& + & \left.\frac{1}{4}\left(\frac{1}{\eta^{2}}+\eta^{2}\right) +\frac{g}{%
\eta}|f|^{2}\right]f+\Gamma f^{},  \label{static_f}
\end{eqnarray}

\begin{equation}
\eta^{4}=1+g|f|^{2}\eta.  \label{eta}
\end{equation}

The energy (Hamiltonian) of the 2D system is (in the most general case,
without assuming relation $f_{1}=f_{2}^{\ast }$ and $\eta _{1}=\eta _{2}$) $%
H=\int \int dxdy\mathcal{H}$, with density
\begin{gather}
\mathcal{H}=\sum_{k=1}^{2}\left\{ \frac{1}{2}|\nabla
f_{k}|^{2}+V(x,y)|f_{k}|^{2}\right.   \notag \\
+(-1)^{k-1}\frac{i}{2}\gamma \left[ f_{k}^{\ast }\left( f_{k}\right)
_{x}-f_{k}\left( f_{k}^{\ast }\right) _{x}\right]   \notag \\
+\left. \frac{1}{4}\left( \frac{1}{\eta _{k}^{2}}+\eta _{k}^{2}\right)
|f_{k}|^{2}+\frac{g}{2\eta _{k}}|f_{k}|^{4}+\Gamma f_{k}f_{3-k}^{\ast
}\right\} .  \label{Hk}
\end{gather}%
In the case when restrictions $f_{1}=f_{2}^{\ast }$ and $\eta _{1}=\eta _{2}$
are imposed, Hamiltonian (\ref{Hk}) simplifies to
\begin{eqnarray}
\mathcal{H} &=&|\nabla f|^{2}+2V(x,y)|f|^{2}+i\gamma \left[ f^{\ast }\left(
f\right) _{x}-f\left( f^{\ast }\right) _{x}\right]   \notag \\
&+&\frac{1}{2}\left( \frac{1}{\eta ^{2}}+\eta ^{2}\right) |f|^{2}+\frac{g}{%
\eta }|f|^{4}+\Gamma \left[ f^{2}+(f^{\ast })^{2}\right] .  \label{H_static}
\end{eqnarray}

Exact solutions to Eq. (\ref{eta}) may be given by the Cardano formula,
\begin{gather}
\eta=\pm\frac{1}{2}\sqrt{\frac{A^{2}-12}{3A}}  \notag \\
+\frac{1}{2}\sqrt{-\frac{A^{2}-12}{3A}\pm2g|f|^{2}\left(\frac{A^{2}-12}{3A}
\right)^{-1/2}},  \label{eta_sol}
\end{gather}
where the upper and lower signs correspond, respectively, to $g>0$ and $g<0$%
, and
\begin{equation}
A\equiv(3/2)^{1/3}\left(9g^{2}|f|^{4}+\sqrt{3}\sqrt{256+27g^{4}|f|^{8}}
\right)^{1/3}.
\end{equation}

Inserting Eq. (\ref{eta_sol}) into Eq. (\ref{static_f}) one gets a 2D NPSE,
which is a generalization of that introduced earlier for the study of
solitons and solitary vortices in ``pancake''-shaped Bose-Einstein
condensates \cite{SalasnichLP02,SalasnichPRA09}.

\subsection{Simple analytical approximations}

Following the consideration of the 1D counterpart of this system in Ref.
\cite{SalasnichPRA13}, a simple analytical approximation can be obtained
from Eq. (\ref{static_f}) for large $\Gamma$:
\begin{equation}
f(x,y;\mu)\approx\left[1+\frac{i\gamma}{\Gamma}\partial_{x}\right]
f_{0}(x,y;\mu^{\prime}),\;\mu^{\prime}\equiv\mu-\Gamma,  \label{b1}
\end{equation}
where $f_{0}(x,y)$ is a real solution obtained from Eqs. (\ref{static_f})
and (\ref{eta}) with $\Gamma=\gamma=0$ and $\mu$ replaced by $\mu^{\prime}$.
Thus, at large $\Gamma$ the solution acquires a small imaginary part given
by Eq. (\ref{b1}).

The term $\sim\gamma$ can be eliminated from Eq. (\ref{static_f}) by means
of substitution \cite{SalasnichPRA13}
\begin{equation}
f(x,y;\mu)=e^{i\gamma x}\widetilde{f}(x,y;\widetilde{\mu}),\;\widetilde{\mu}
\equiv\mu-\gamma^{2}/4,  \label{L_gamma}
\end{equation}
which transforms Eqs. (\ref{static_f}) and (\ref{eta}) into the following
form:
\begin{eqnarray}
\widetilde{\mu}\widetilde{f} & = & \left[-\frac{1}{2}(\partial_{x}^{2}
+\partial_{y}^{2})+V(x,y)\right.  \notag \\
& + & \left.\frac{1}{4}\left(\frac{1}{\widetilde{\eta}^{2}} +\widetilde{\eta}%
^{2}\right)+\frac{g}{\widetilde{\eta}} |\widetilde{f}|^{2}\right]\widetilde{f%
} +\Gamma e^{-2i\gamma x}\widetilde{f}^{},  \label{static_f-1}
\end{eqnarray}

\begin{equation}
\widetilde{\eta}^{4}=1+g|\widetilde{f}|^{2}\widetilde{\eta}.  \label{eta-1}
\end{equation}
Thus, in the absence of the Rabi term ($\Gamma =0$), the SO coupling drops
from Eq. (\ref{static_f-1}). Again following the lines of Ref. \cite%
{SalasnichPRA13}, an approximate solution to Eq. (\ref{static_f-1}) can be
constructed for the limit of large $\gamma$:
\begin{equation}
\widetilde{f}(x,y;\mu)\approx\left[1-\frac{\Gamma}{\gamma^{2}}e^{-2i\gamma
x} \right]\widetilde{f}_{0}(x,y;\widetilde{\mu}),  \label{b2}
\end{equation}
where $\widetilde{f}_{0}(x,y;\widetilde{\mu})$ stands for the usual real
solution to Eqs. (\ref{static_f-1}) and (\ref{eta-1}) with $\widetilde{\mu}$
taken as per Eq. (\ref{L_gamma}).

\subsection{Low- and high-density limits}

In the low-density limit, $|g||f|^{2}\ll1$, Eq. (\ref{eta}) yields a simple
solution,
\begin{equation}
\eta_{\mathrm{low}}\simeq1+g|f|^{2}/4.  \label{eta_low}
\end{equation}
The substitution of this approximation in Eq. (\ref{static_f}) reduces it to
the NPSE with the cubic-quintic nonlinearity, in which the quintic term
always corresponds to the effective self-attraction:
\begin{eqnarray}
\left(\mu-\frac{1}{2}\right)f & = & \left[-\frac{1}{2}(\partial_{x}^{2}
+\partial_{y}^{2})+V(x,y)+i\gamma\partial_{x}\right.  \notag \\
& + & \left.g|f|^{2}-\frac{3}{16}g^{2}|f|^{4}\right]f+\Gamma f^{\ast}.
\label{CQ}
\end{eqnarray}

In the high-density limit, $|g||f|^{2}\gg1$, asymptotic expressions for the
axial width which follow from Eq. (\ref{eta}) are different in the cases of
repulsion ($g>0$) and attraction ($g<0$),
\begin{equation}
\eta_{\mathrm{high}}^{(\mathrm{rep})}\simeq(g|f|^{2})^{1/3}, \quad\eta_{%
\mathrm{high}}^{(\mathrm{attr})}\simeq-(g|f|^{2})^{-1}.  \label{eta_high}
\end{equation}
The substitution of these approximations in Eq. (\ref{static_f}) leads to
two different 2D NPSEs: in the case of repulsion, it is
\begin{eqnarray}
\mu f & = & \left[-\frac{1}{2}(\partial_{x}^{2}+\partial_{y}^{2})+V(x,y)
+i\gamma\partial_{x}\right.  \notag \\
& + & \left.\frac{5}{4}g^{2/3}|f|^{4/3}\right]f+\Gamma f^{\ast},  \label{4/3}
\end{eqnarray}
and in the case of attraction, the effective equation is the NPSE with the
quintic term only:

\begin{eqnarray}
\mu f & = & \left[-\frac{1}{2}(\partial_{x}^{2}+\partial_{y}^{2})+V(x,y)
+i\gamma\partial_{x}\right.  \notag \\
& - & \left.\frac{3}{4}g^{2}|f|^{4}\right]f+\Gamma f^{}.  \label{quintic}
\end{eqnarray}

\begin{figure}[tbp]
\centering
\includegraphics[width=0.9\columnwidth]{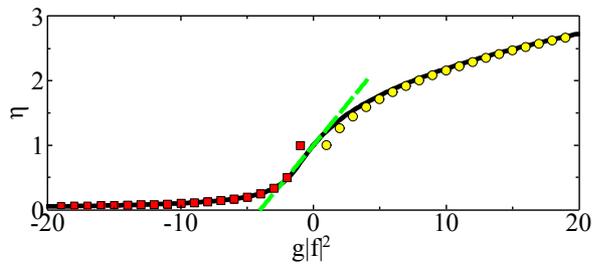}
\caption{(Color online) The axial width $\protect\eta$ of the pancake-shaped
condensate versus $g|f|^{2}$. The solid (black) curve displays the behavior
of $\protect\eta$ given by the Eq. (\protect\ref{eta_sol}); by the dashed
(green) line we show the low-density limit {[}Eq. (\protect\ref{eta_low}){]}%
; the high-density limit for the repulsive and attractive cases {[}Eq. (%
\protect\ref{eta_high}){]} are shown by circles (yellow) and boxes (red),
respectively.}
\label{F1}
\end{figure}

Figure \ref{F1} shows the behavior of the axial width $\eta$ as a function
of $g|f|^{2}$ for the different cases presented above. The exact solution
for $\eta$ (given by Eq. (\ref{eta_sol})) is displayed by the solid (black)
line. The solution for the low-density limit {[}Eqs. (\ref{eta_low}){]} is
presented by the dashed (green) line, while circles (yellow) and boxes (red)
represents the high-density repulsive and attractive limits, respectively,
see Eq. (\ref{eta_high}).

\section{Numerical Results}

We start by present the numerical results for the effective 2D stationary
equation (\ref{static_f}). To this end, we have employed a split-step
algorithm that uses the imaginary-time propagation (ITP) to generate the
ground-state solution of the system. The ITP method includes the restoration
the original norm (\ref{norma})\ of the solution at the end of each step of
marching forward in imaginary time. To this end, Eq. (\ref{static_f}) was
replaced by
\begin{equation}
iF_{t}=(H_{d}+H_{nd})F+\Gamma F^{\ast },  \label{a1}
\end{equation}%
where
\begin{equation}
H_{d}\equiv -\frac{1}{2}\left( \partial _{x}^{2}+\partial _{y}^{2}\right)
+i\gamma \partial _{x},  \label{a2}
\end{equation}%
\begin{equation}
H_{nd}\equiv V(x,y)+G(|F|^{2}),  \label{a3}
\end{equation}%
and $F\equiv F_{1}=F_{2}^{\ast }$ {[}cf. Eqs. (\ref{Fk}) and (\ref{Fk2}){]},
the subscripts $d$ and $nd$ referring to the derivative and non-derivative
terms, respectively, and $G(|F|^{2})$ is the nonlinear term of the equation
that we aim to solve. For the implementation of the ITP method, we change $%
t\rightarrow -it$ in Eq. (\ref{a1}). Then, we take advantage of the
Baker-Campbell-Hausdorff formula, and omit terms $\sim $ $\Delta t^{2}$ to
formally integrate Eq. (\ref{a1}) term by term. The derivative operator (\ref%
{a2}) was handled by means of the Crank-Nicholson algorithm, and the
Runge-Kutta algorithm was used in the last term of Eq. (\ref{a1}). We
employed the spatial and temporal steps $\Delta x=\Delta y=0.1$ and $\Delta
t=0.01$, respectively. The input was chosen as $\psi (x,y,0)=\pi ^{-1/4}\exp
\left( -x^{2}/2\right) $, and the output was picked up when the convergence
of the energy attained the level of $10^{-8}$ (or after $10^{4}$ iterations,
in the case of unstable solutions).

The stability of the solutions obtained by the ITP method was checked by
subsequent real-time simulations (RTS) of Eq. (\ref{full_2D}) for the
evolution of perturbed stationary profiles, to which we have added random
perturbations at the $5\%$ amplitude level. A detailed discussion of these
numerical methods is given in Ref. \cite{MuruganandamCPC09}.

\subsection{Repulsive interatomic interactions}

Here, we consider the BEC with the repulsive nonlinearity ($g>0$), trapped
in HO potential
\begin{equation}
V(x,y)=\frac{\lambda^{2}}{2}(x^{2}+y^{2}),  \label{POT}
\end{equation}
with $\lambda\equiv\omega_{\bot}/\omega_{x}\ll1$ representing the anisotropy
of the HO confinement.

\begin{figure}[tbp]
\centering
\includegraphics[width=0.3\columnwidth]{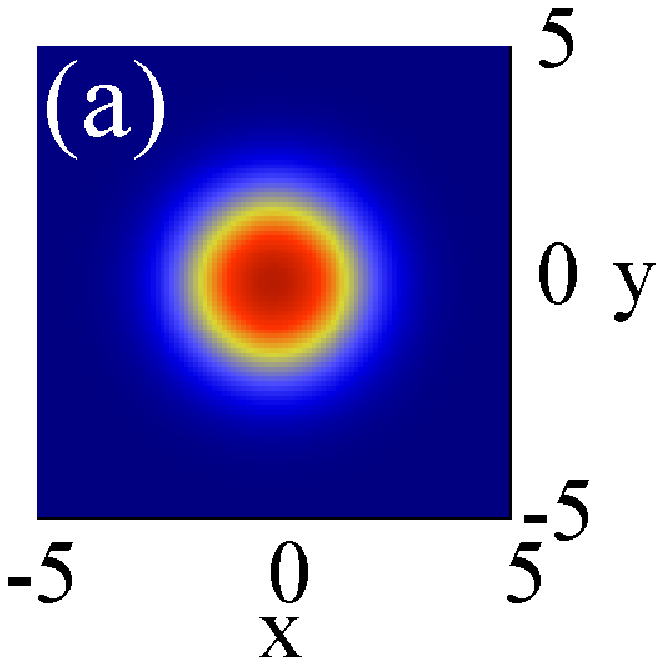} \hfil
\includegraphics[width=0.3\columnwidth]{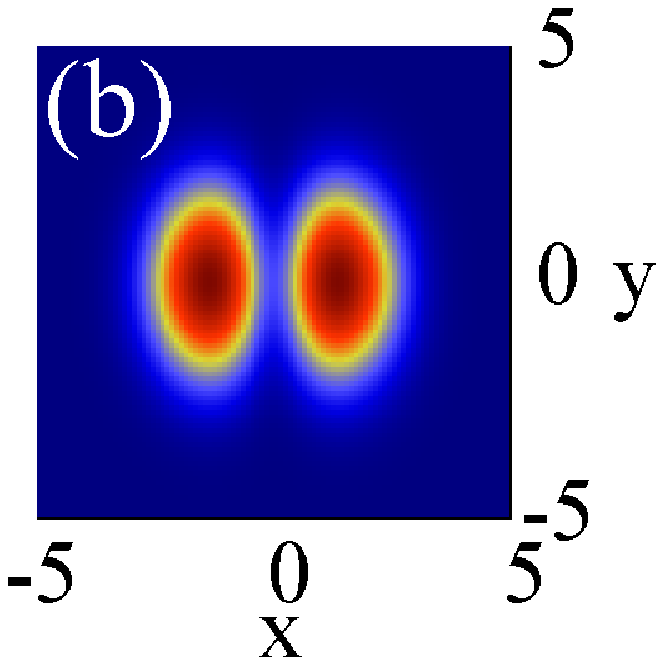} \hfil %
\includegraphics[width=0.3\columnwidth]{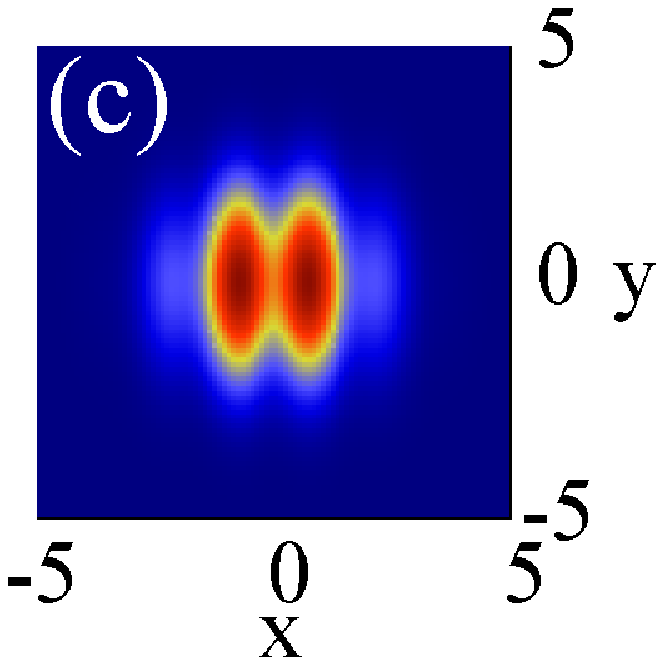} \hfil
\includegraphics[width=0.3\columnwidth]{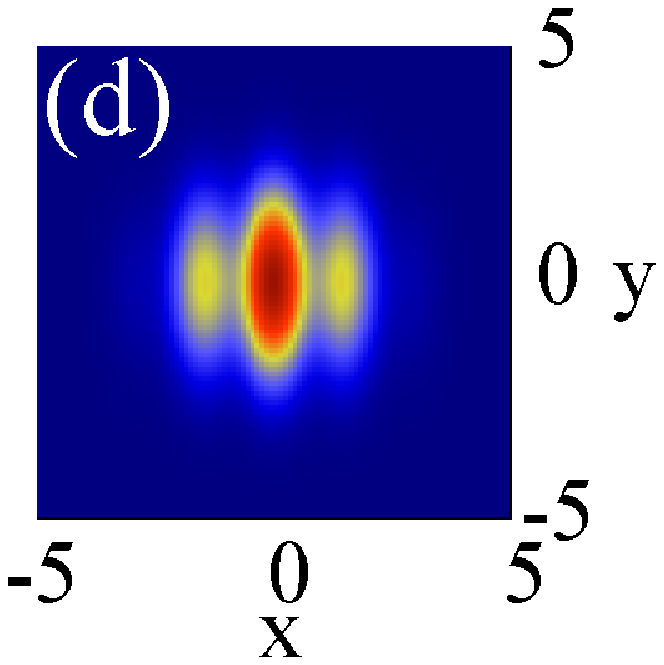} \hfil %
\includegraphics[width=0.3\columnwidth]{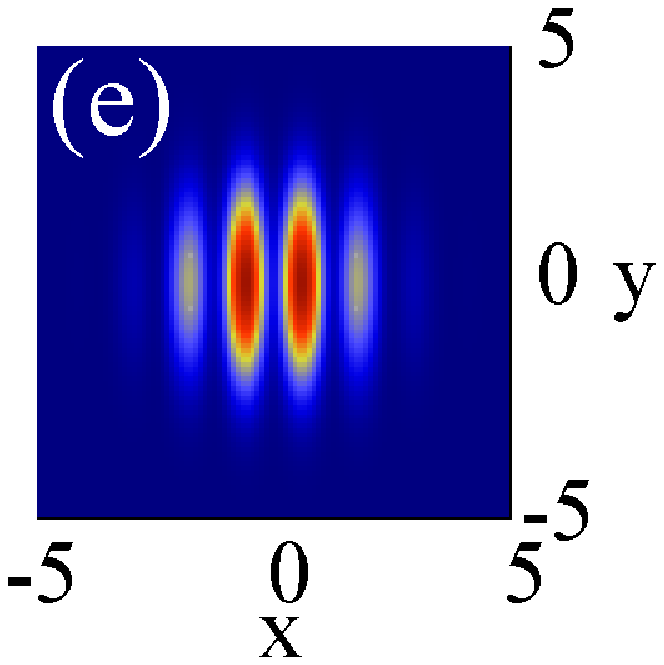} \hfil
\includegraphics[width=0.3\columnwidth]{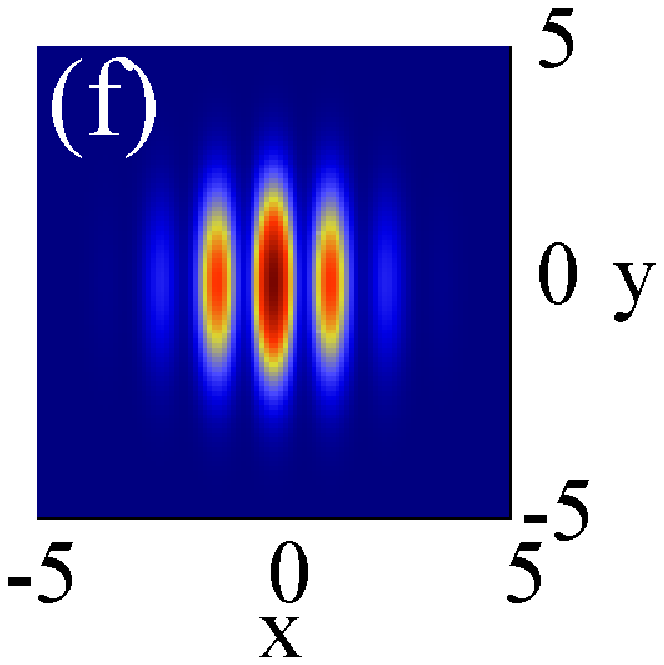}
\caption{(Color online) Density profile $|f|^{2}$ produced by Eq. (\protect
\ref{static_f}) for (a) $\protect\gamma=1$ and $\Gamma=0$; (b) $\protect%
\gamma=1$ and $\Gamma=1$; (c) $\protect\gamma=2$ and $\Gamma=1$; (d) $%
\protect\gamma=2$ and $\Gamma=-1$; (e) $\protect\gamma=3$ and $\Gamma=5$;
(f) $\protect\gamma=3$ and $\Gamma=-5$; we use $g=1$ ($\mathrm{max}\left[%
g|f|^{2}\right]\simeq0.1$) and $\protect\lambda=1/3$ in all cases. }
\label{F2}
\end{figure}

\begin{figure}[tbp]
\centering
\includegraphics[width=0.3\columnwidth]{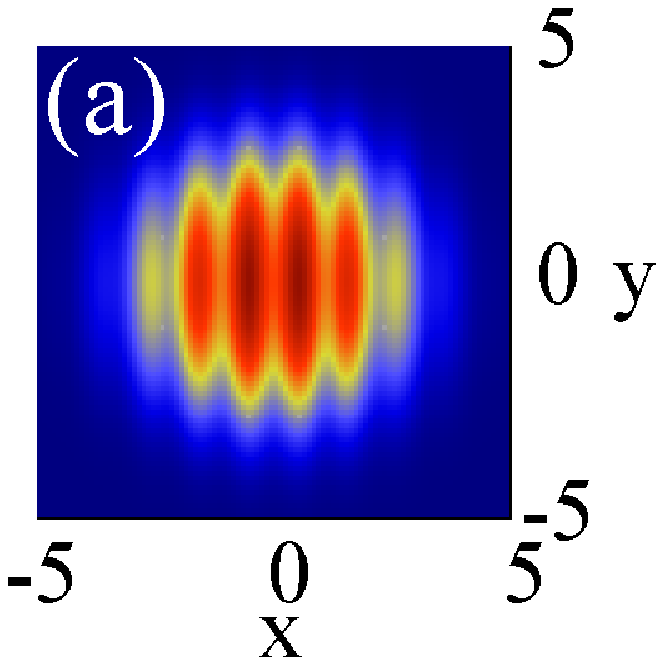} \hfil
\includegraphics[width=0.3\columnwidth]{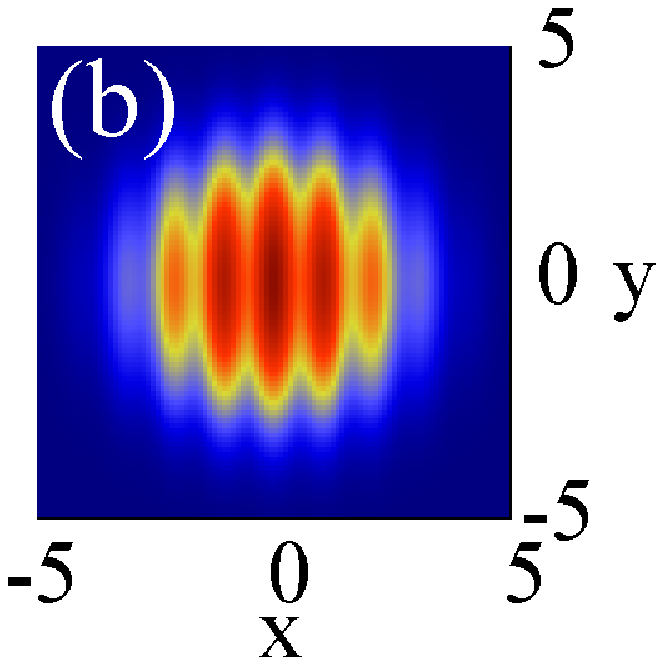} \includegraphics[width=0.7%
\columnwidth]{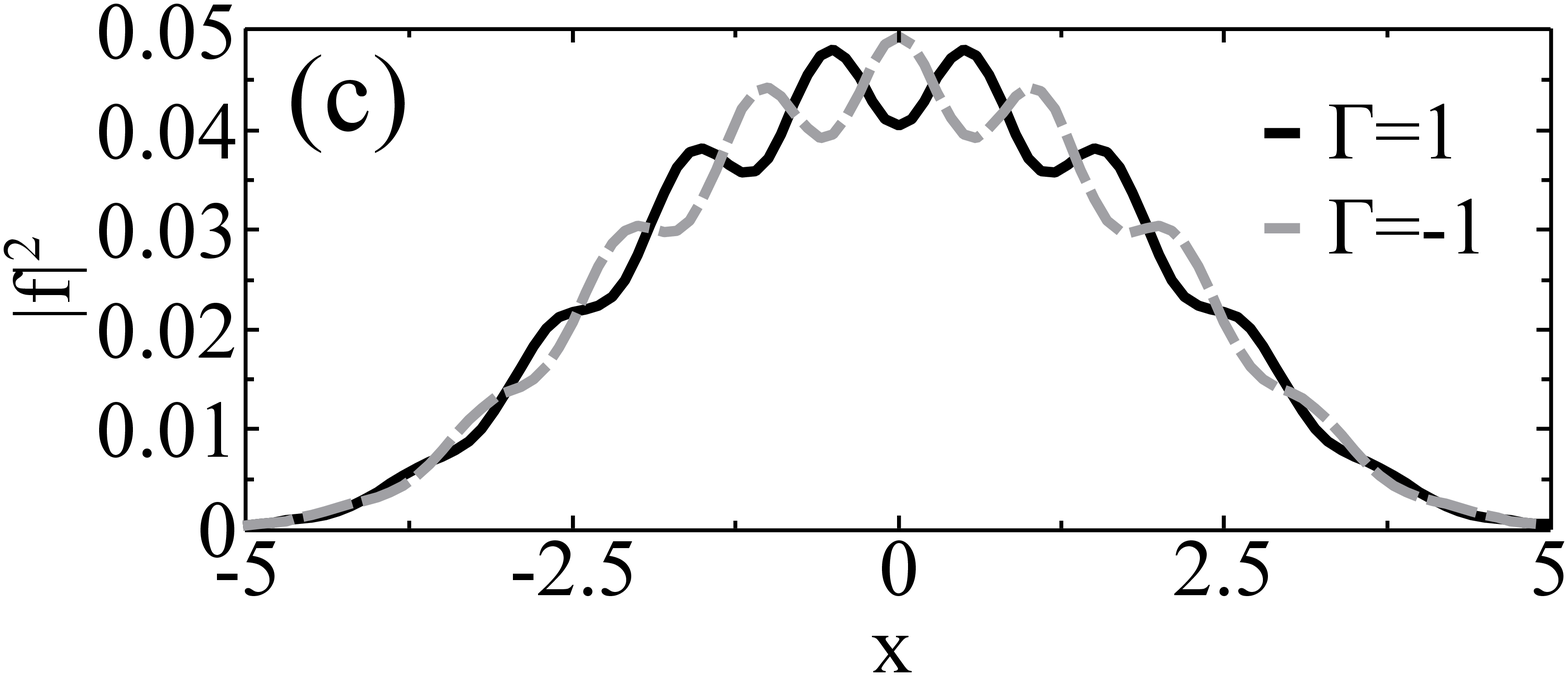}
\caption{(Color online) Density profile $|f|^{2}$ produced by Eq. (\protect
\ref{static_f}) for (a) $\Gamma=1$ and (b) $\Gamma=-1$. (c) Cross-section
profiles in $x$-direction, $|f(x,0)|^{2}$, corresponding to cases (a) (the
solid black line) and (b) (the dashed gray line). (c) In both cases we have
used $g=20$ ($\mathrm{max}\left[g|f|^{2}\right]\simeq1$), $\protect\lambda%
=1/3$, and $\protect\gamma=3$.}
\label{F3}
\end{figure}

First, we set $g=1$, to address the low-density case, $g|f|^{2}\ll1$. In
Fig. \ref{F2} we display the numerical results for different values of the
SO and Rabi constants, $\gamma$ and $\Gamma$. As shown in Fig. \ref{F2}(a),
for $\Gamma=0$ the ground state does not feature an oscillation pattern,
while Fig. \ref{F2}(b) displays the influence of positive $\Gamma$ on the
solution. Note that in Figs. \ref{F2}(c) and \ref{F2}(e), for $\Gamma<\gamma$
and $\Gamma>\gamma$, respectively, the results show the emergence of spatial
oscillations in the solution, with an amplitude approximately proportional
to $\Gamma/\gamma^{2}$ or $\gamma/\Gamma$, cf. Eqs. (\ref{b2}) and (\ref{b1}%
), respectively. We stress the importance of sign of $\Gamma$, which
determines a peak or a hole at the center of the solution, as can be seen in
Figs. \ref{F2}(d) and \ref{F2}(f), demonstrating the inverted sign with
respect to that in Figs. \ref{F2}(c) and \ref{F2}(e), respectively. This
result is also emphasized in the high-density limit, as shown in Fig. \ref%
{F3} for $\Gamma=1$ and $\Gamma=-1$ {[}Figs. \ref{F3}(a) and \ref{F3}(b),
respectively{]}, setting $g=20$, $\lambda=1/3$, and $\gamma=3$ in both
cases. Also, in Fig. \ref{F3}(c) we present, for the sake of comparison, two
cross-section profiles drawn through $y=0$.

\begin{figure}[tbp]
\centering
\includegraphics[width=0.48\columnwidth]{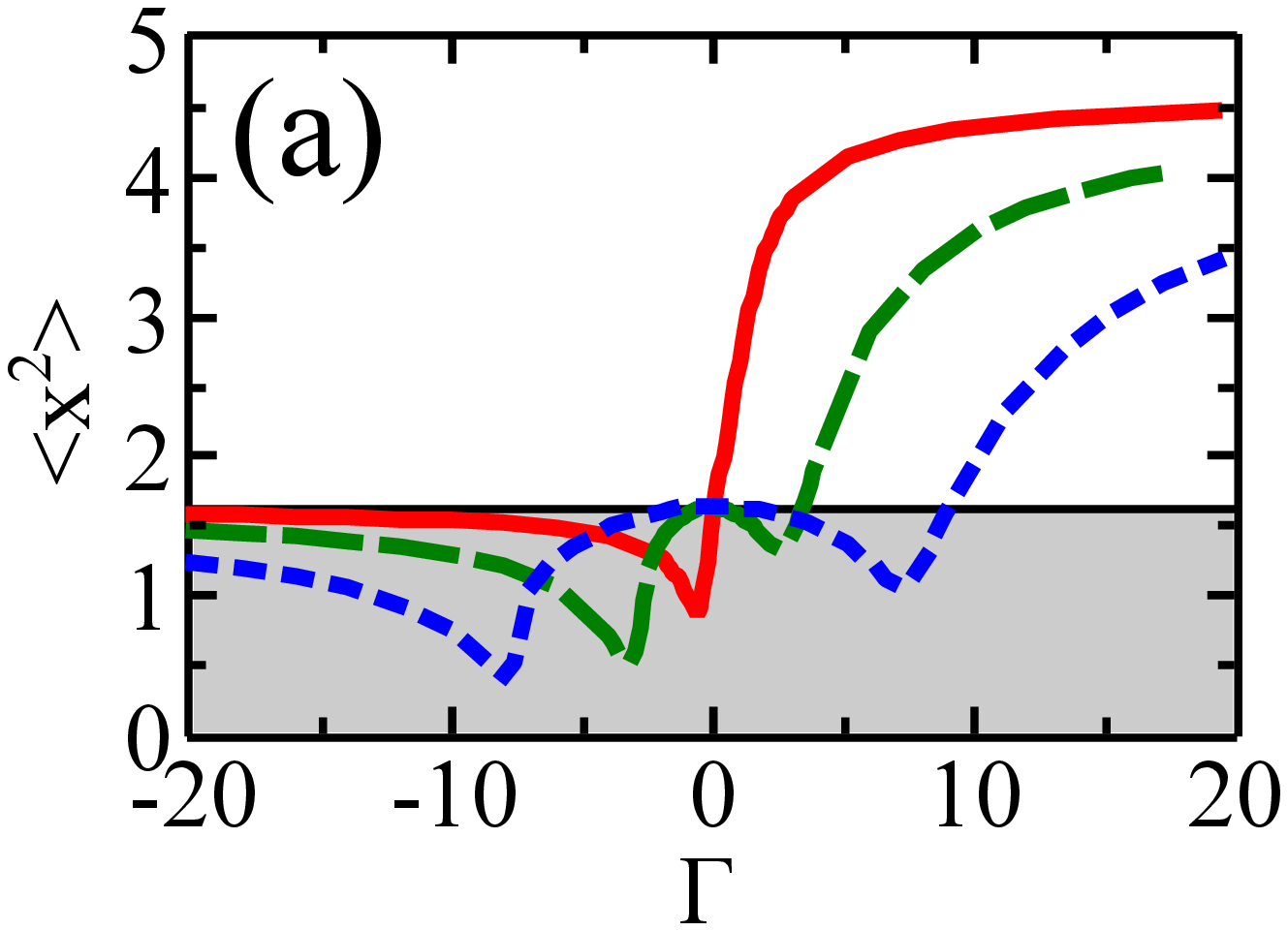} \includegraphics[width=0.48%
\columnwidth]{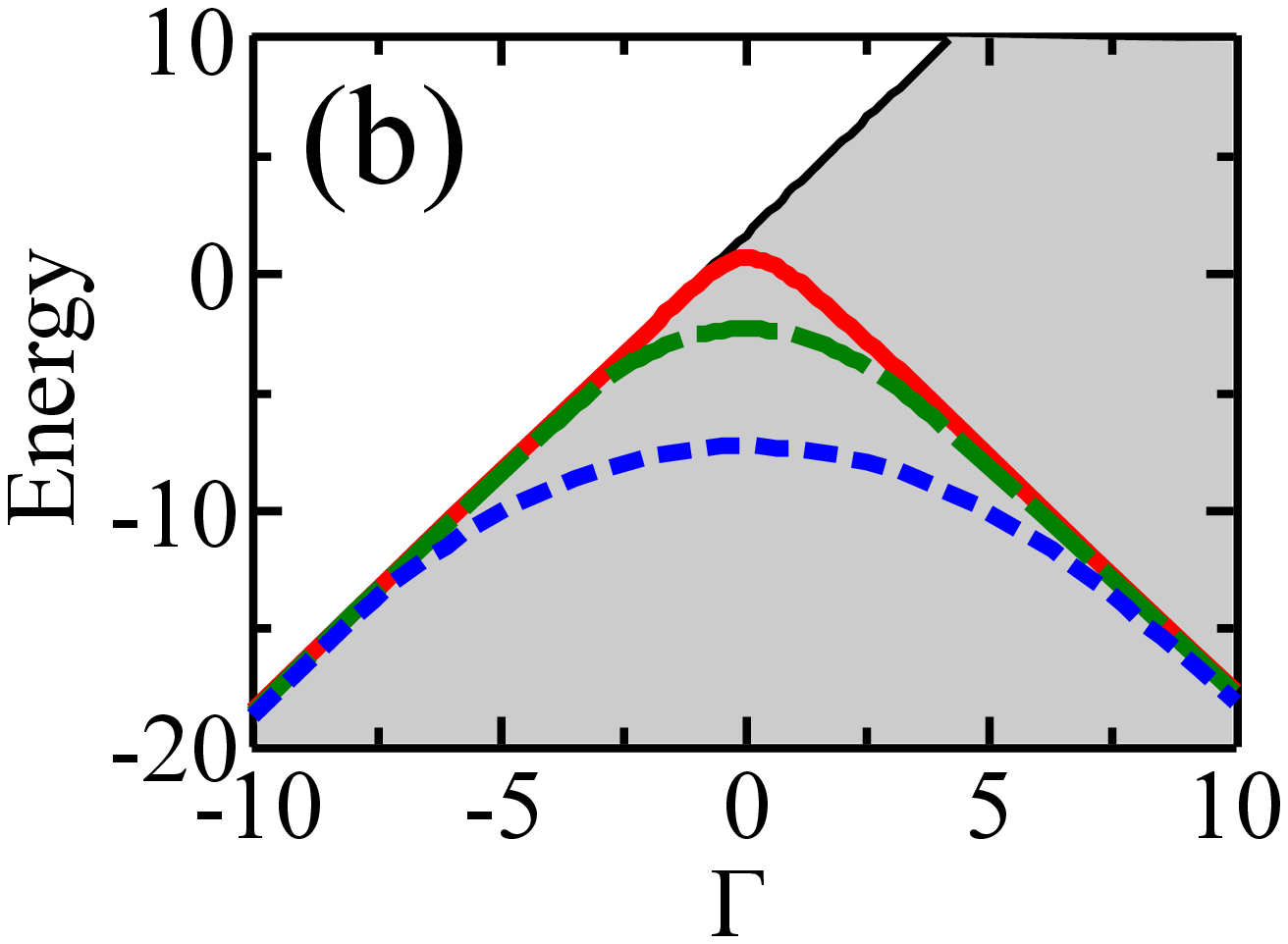}
\caption{(Color online) (a) The squared width, $\left\langle
x^{2}\right\rangle $, and (b) the energy of the solitons versus the Rabi
coupling, $\Gamma$. The boundary of the gray region corresponds to the
respective value for $\protect\gamma=0$, while $\protect\gamma=1$ is
represented by the red solid line, $\protect\gamma=2$ -- by the green
long-dashed-line, and $\protect\gamma=3$ -- by the dashed-line blue line.
The other parameters are $g=1$ and $\protect\lambda=1/3$.}
\label{F4}
\end{figure}

In Fig. \ref{F4}(a) we display the average squared width,
\begin{equation}
\left\langle x^{2}\right\rangle =\int_{-\infty}^{+\infty}
\int_{-\infty}^{+\infty}x^{2}|f|^{2}dxdy,  \label{width}
\end{equation}
versus $\Gamma$ for different values of $\gamma$, with $f$ obtained
numerically from Eq. (\ref{static_f}). Here, as well as in Fig. \ref{F2},
the influence of the sign of $\Gamma$ on the form of the solution is
evident. At $\Gamma>0$ the splitting in the solution profile is such that
the squared width is larger than its value at $\gamma=0$. The corresponding
energies versus $\Gamma$ are shown in Fig. \ref{F4}(b).

\subsection{Attractive interatomic interactions}

In this subsection we deal with the attractive nonlinearity ($g<0$), with
the corresponding sign changes in Eq. (\ref{eta_sol}). First, we aim to
investigate effects of the SO and Rabi couplings on bright solitons by
solving Eqs. (\ref{full_2D}) and (\ref{static_f}) with $g<0$, in the absence
of the axial trapping potential {[}$V(x,y)=0${]}, following the similar
analysis reported in Ref. \cite{SakaguchiPRE14} for the usual cubic
nonlinearity.

For $\gamma =0$ and/or $\Gamma =0$, with $V(x,y)=0$ in Eqs. (\ref{full_2D})
and (\ref{static_f}), all solutions are unstable, as should be expected for
2D free-space solitons in the case of the self-attraction. However, it was
found in Ref. \cite{SakaguchiPRE14} that two-component BEC with the SO
Rashba coupling and cubic attractive interactions gives rise to stable
solitons of two types: semi-vortices (with a vortex in one component and a
fundamental soliton in the other), or mixed modes (with topological charges $%
0$ and $\pm 1$ mixed in both components). Here, we report numerical
observation of a stable composite bright solitons, following a scenario
different from that shown in Ref. \cite{SakaguchiPRE14}: we consider the
Rashba-equal-Dresselhaus coupling, while in Ref. \cite{SakaguchiPRE14} the
Rashba-only SO term was analyzed. Note that, in the case of the attractive
interactions ($g < 0$), our solution does not represent the ground state,
which formally corresponds to the collapse in the 2D setting.

\begin{figure}[tb]
\centering
\includegraphics[width=0.48\columnwidth]{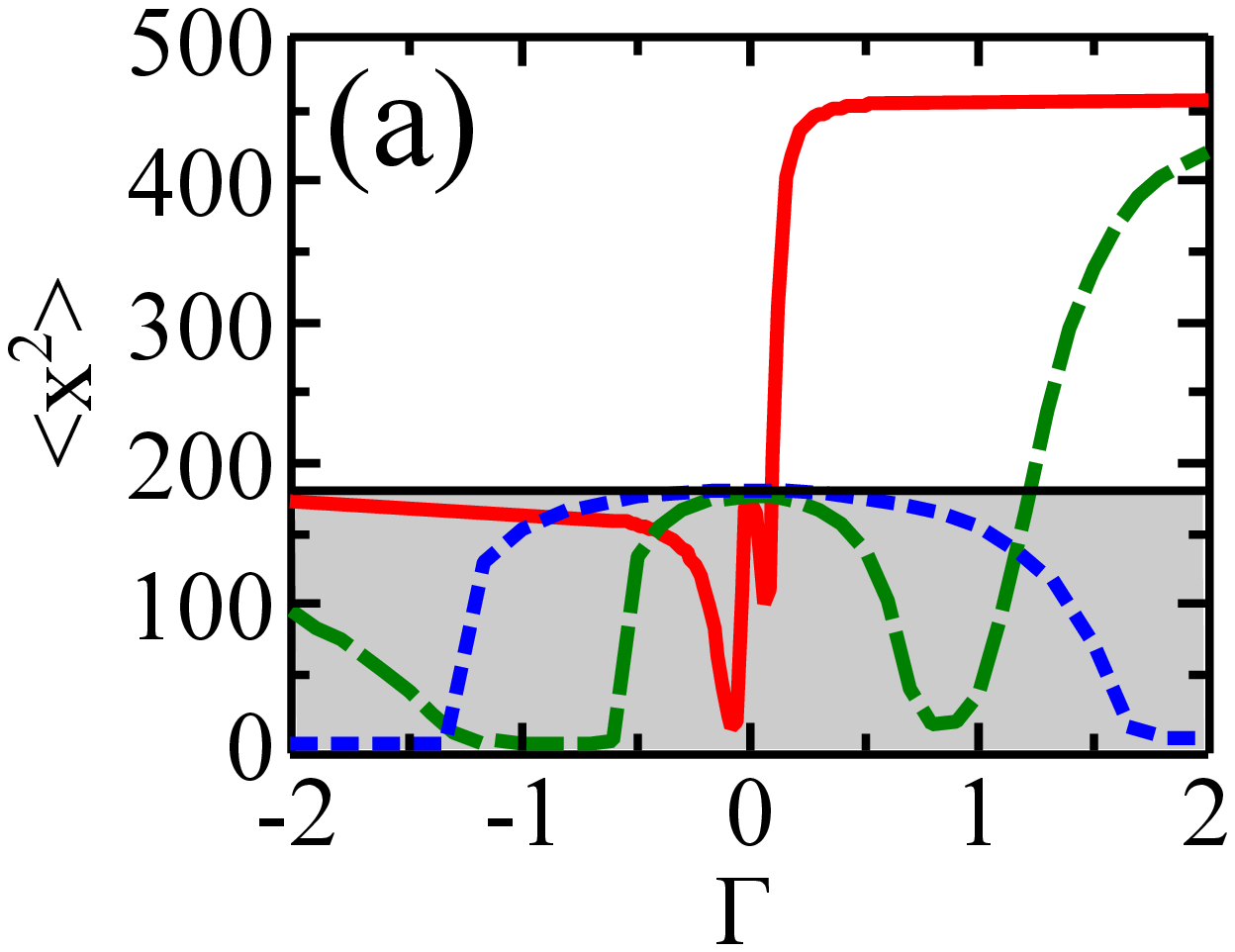} \includegraphics[width=0.48%
\columnwidth]{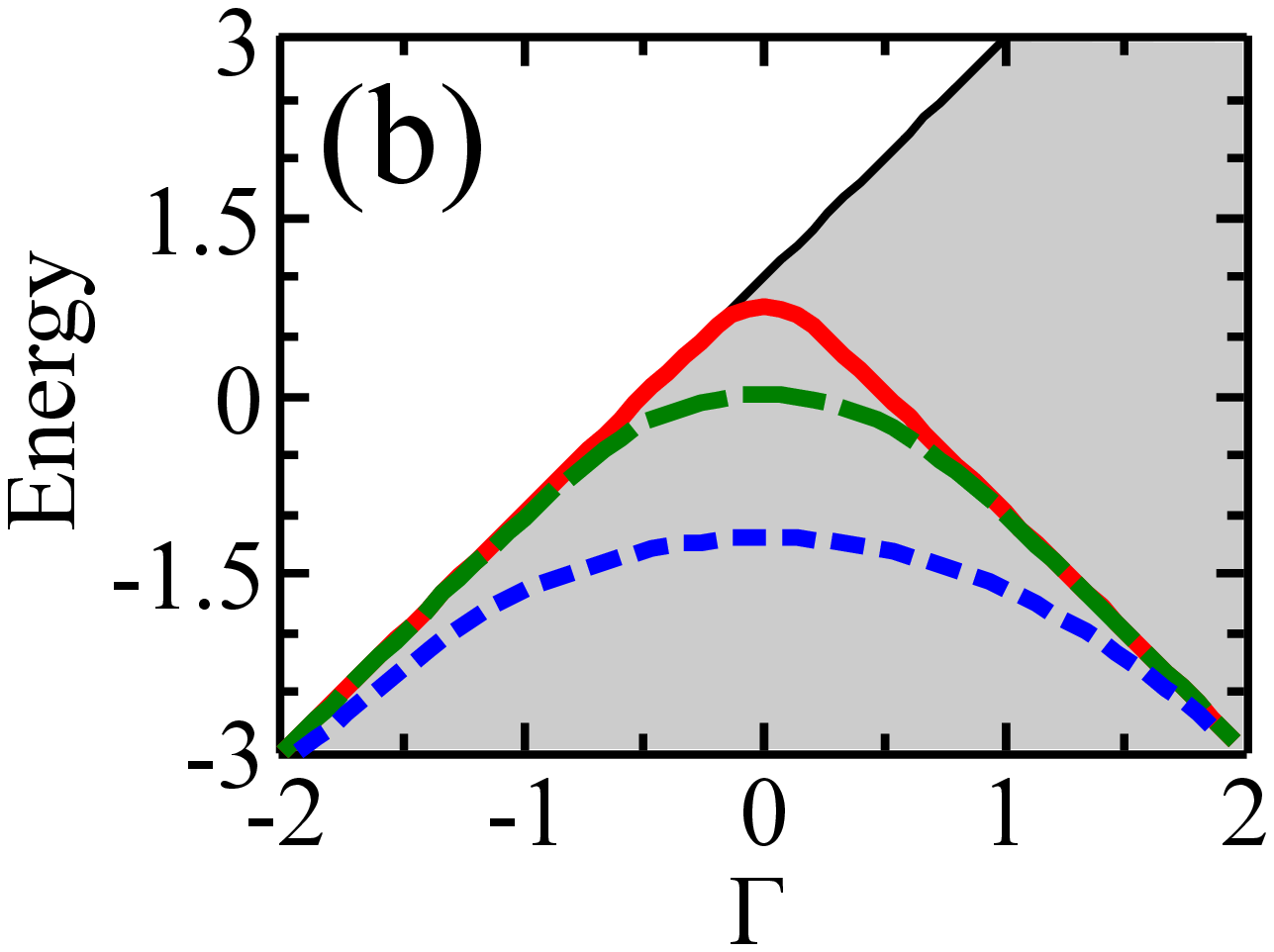}
\caption{(Color online) (a) The squared width, $\left\langle
x^{2}\right\rangle $, and (b) the energy of the solitons versus the Rabi
coupling $\Gamma$ for BEC with the attractive interactions ($g=-3$). The
boundary of the gray region corresponds to $\protect\gamma=0$, while $%
\protect\gamma=0.5$ corresponds to the red solid line, $\protect\gamma=1$ --
to the green long-dashed-line, and $\protect\gamma=1.5$ -- to the blue
dashed line. We use $\protect\lambda=0$ (free space).}
\label{F5}
\end{figure}

To highlight the possible stability regions for 2D solitons, we look first
for solutions with a small width, measured as per definition (\ref{width}).
The squared width, $\left\langle x^{2}\right\rangle $, and the energy for
states supported by the attractive interaction are displayed in Fig. \ref{F5}%
(a) and \ref{F5}(b), respectively, as a function of $\Gamma$ for $\gamma=0.5$
by the red solid line, for $\gamma=1$ by the green long-dashed-line, and for
$\gamma=1.5$ by the blue dashed line. The boundary of the gray region shows
the values in the absence of the SO coupling ($\gamma=0$). Note that the
squared width is more sensitive to the variation of $\Gamma$ in comparison
to the case of the repulsive interaction, cf. Fig. \ref{F4}(a). However, the
energy of the solutions shows a pattern similar to that observed in the
repulsive system, cf. Fig. \ref{F4}(b). Actually, minima of the squared
width are linked to the stability regions, see below.

\begin{figure}[tb]
\centering
\includegraphics[width=0.48\columnwidth]{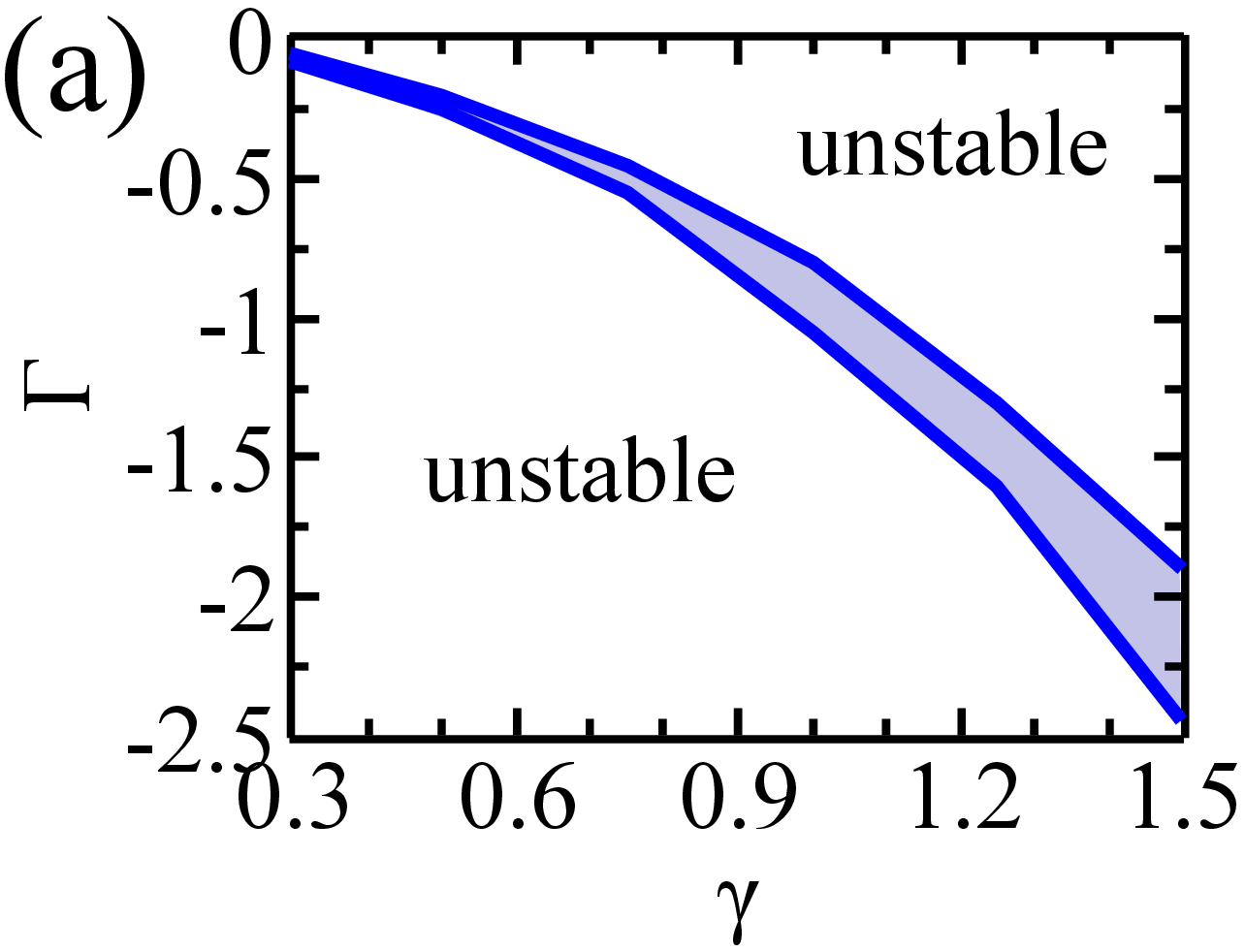} \includegraphics[width=0.48%
\columnwidth]{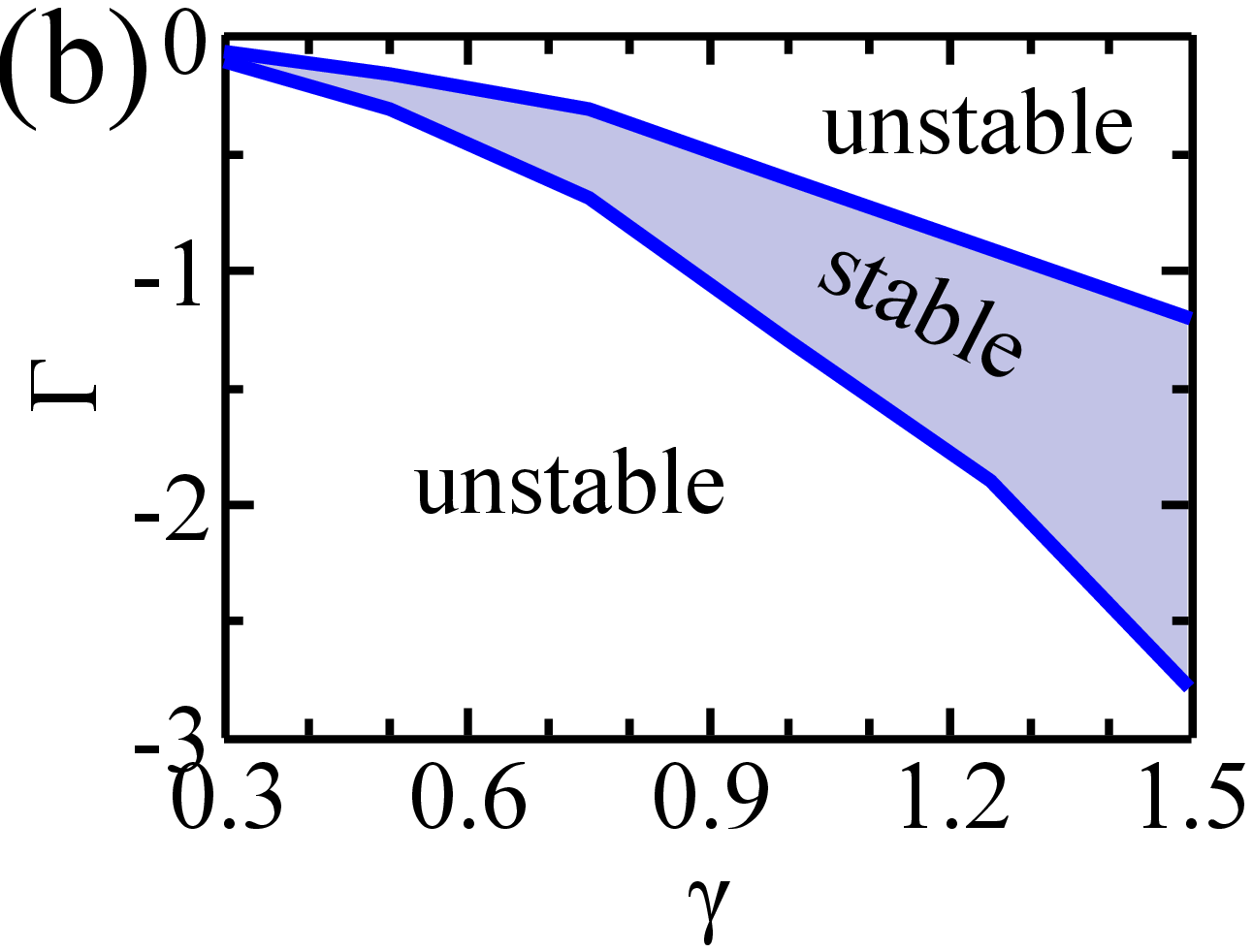}
\caption{(Color online) Stability region for the ground-state soliton
solutions, as a function of the SO and Rabi coupling strengths. We set $g=-1$
in (a) and $g=-3$ in (b), both with $\protect\lambda=0$ (free space). Stable
solutions are found in the highlighted regions for $\Gamma<0$.}
\label{F6}
\end{figure}

In Fig. \ref{F6} we show the main result of this subsection, \textit{viz}.,
\emph{stability regions} for the 2D free-space solitons for (a) $g=-1$ and
(b) $g=-3$, as produced by direct RTS of the perturbed evolution of
stationary solutions that were obtained by means of the ITP method. For $%
\Gamma <0$ we find a region of stable single-peak solutions, while for $%
\Gamma >0$ all the solutions are unstable (at least, for $\Gamma <2$). Note
that the stability region slightly increases with the increase of the
strength of the nonlinearity.

\begin{figure}[tb]
\centering \includegraphics[width=0.3\columnwidth]{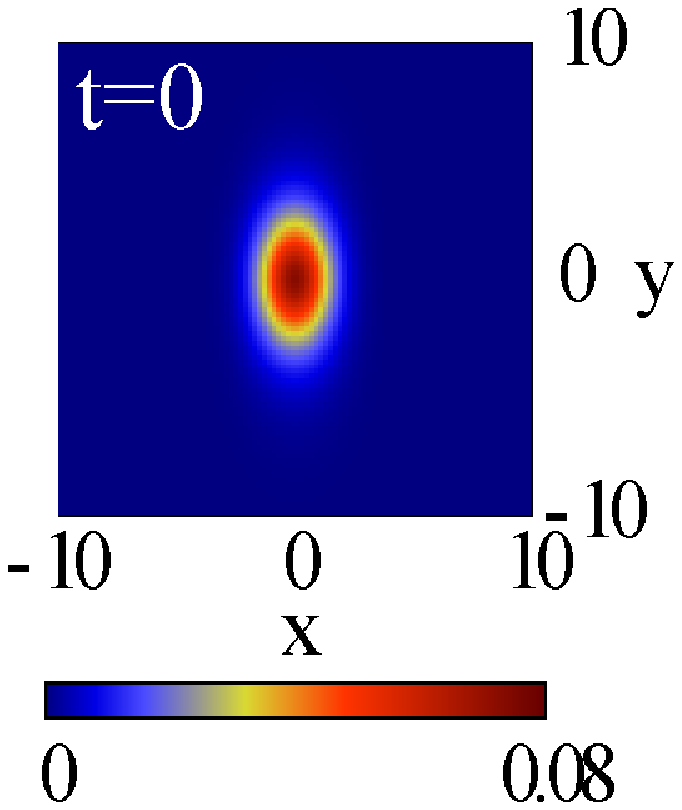} %
\includegraphics[width=0.3\columnwidth]{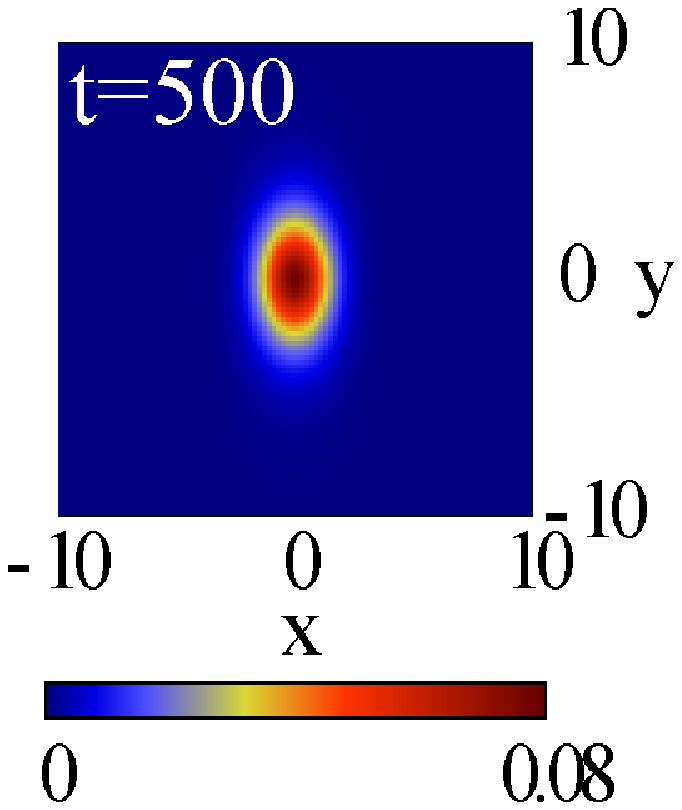} \includegraphics[width=0.3%
\columnwidth]{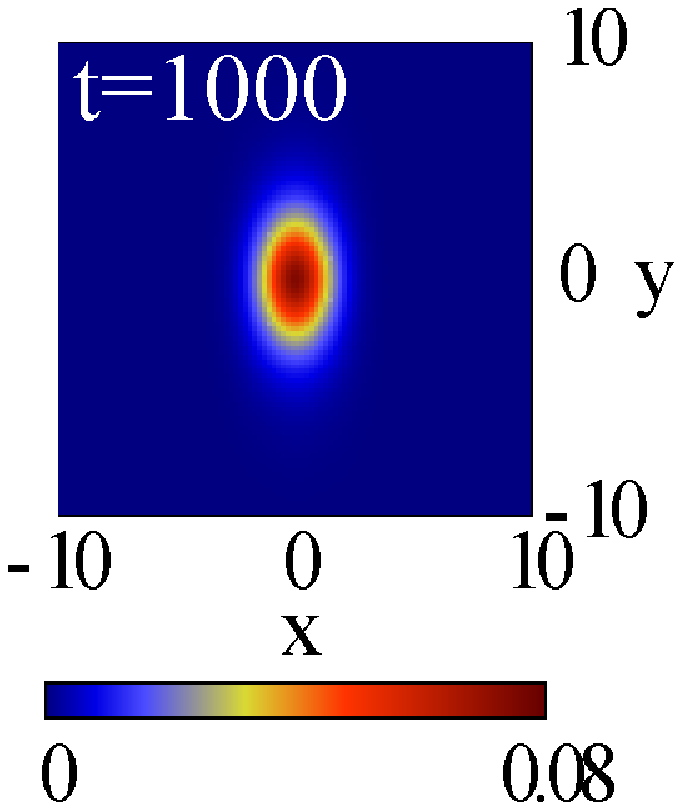}
\caption{(Color online) The stable evolution in the free space ($\protect%
\lambda =0$) of the perturbed solution, $|F_{1}|^{2}$, whose stationary form
was produced by the ITP method for $g=-3$, $\protect\gamma =1$, and $\Gamma
=-1$ [note that this point belongs to the stability region in Fig. \protect
\ref{F6}(b)]. A similar result (not shown here) is obtained for the other
component, $|F_{2}|^{2}$.}
\label{F8}
\end{figure}

\begin{figure}[tb]
\centering
\includegraphics[width=0.3\columnwidth]{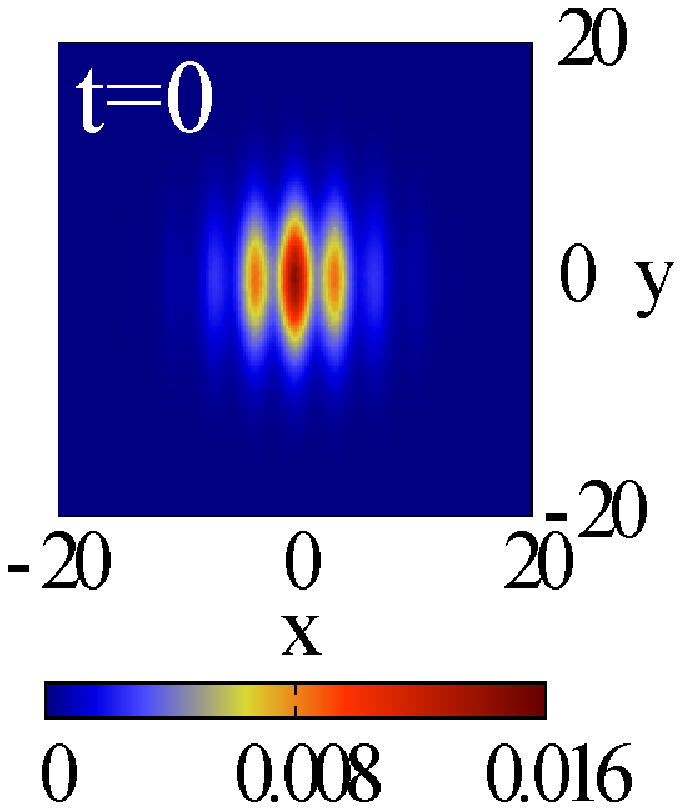} \includegraphics[width=0.3%
\columnwidth]{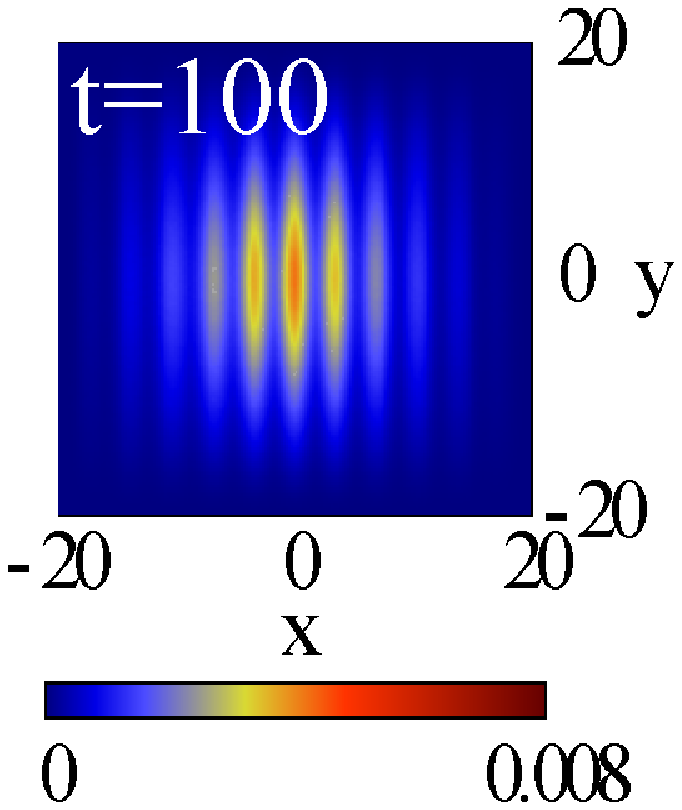} \includegraphics[width=0.3\columnwidth]{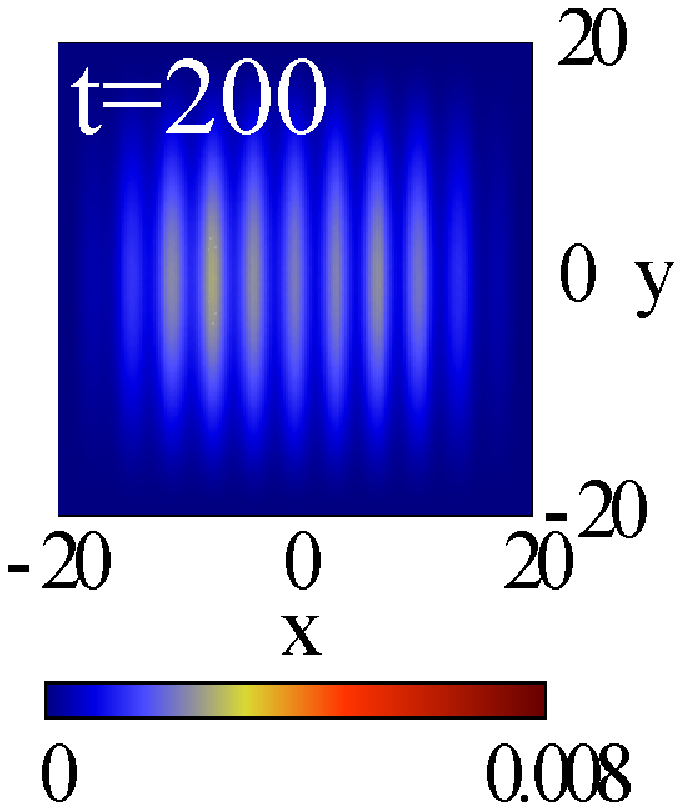}
\caption{(Color online) The unstable evolution in the free space ($\protect%
\lambda =0$) of the perturbed solution, $|F_{1}|^{2}$, whose stationary form
was produced by the ITP method for $g=-3$, $\protect\gamma =1$, and $\Gamma
=-0.4$ [note that this point belongs to the instability region in \protect
\ref{F6}(b)]. A similar result (not shown here) is obtained for the other
component, $|F_{2}|^{2}$.}
\label{F9}
\end{figure}

As might be expected, the stability of the solutions obtained by the ITP
method has been corroborated by subsequent real-time simulations (RTS) of
Eq. (\ref{full_2D}). In Figs. \ref{F8} and \ref{F9}, respectively, we show
examples of the stable and unstable real-time evolution of perturbed
stationary solutions that were supplied by the ITP method. All stable
solutions that we have found are fundamental solitons (in contrast to stable
semi-vortices and mixed modes obtained in Ref. \cite{SakaguchiPRE14}), with
a single-peak shape similar to that observed in Fig. \ref{F8}.

\begin{figure}[tb]
\centering
\includegraphics[width=0.3\columnwidth]{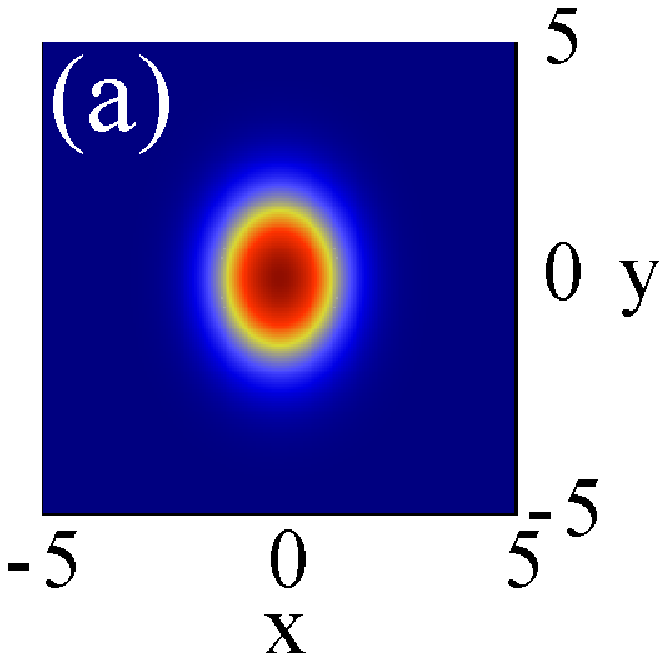} \hfil
\includegraphics[width=0.3\columnwidth]{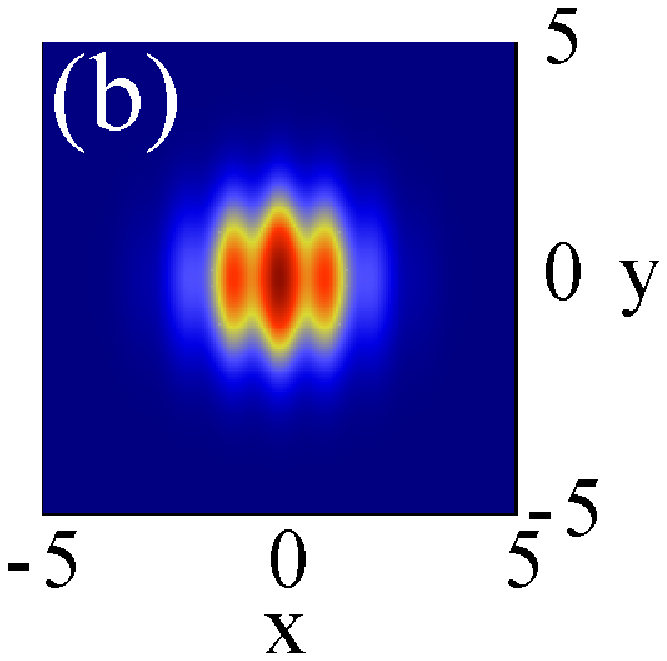} \hfil
\includegraphics[width=0.3\columnwidth]{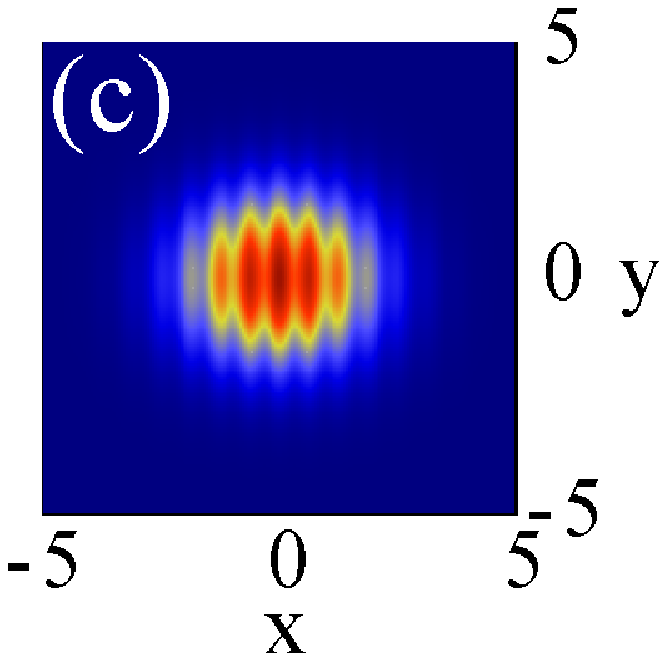} \hfil
\includegraphics[width=0.3\columnwidth]{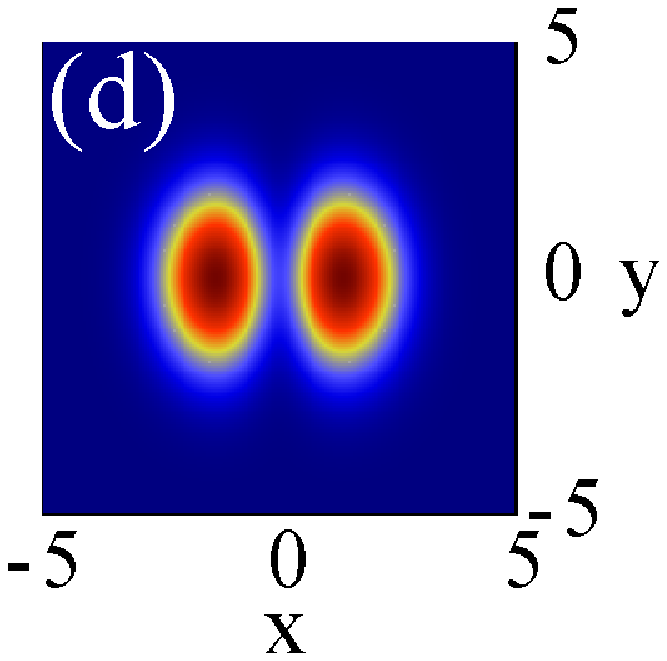} \hfil
\includegraphics[width=0.3\columnwidth]{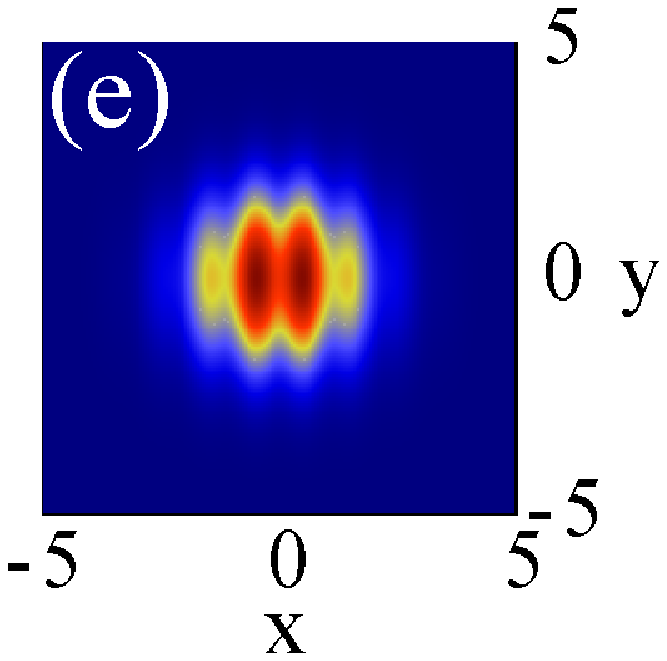} \hfil
\includegraphics[width=0.3\columnwidth]{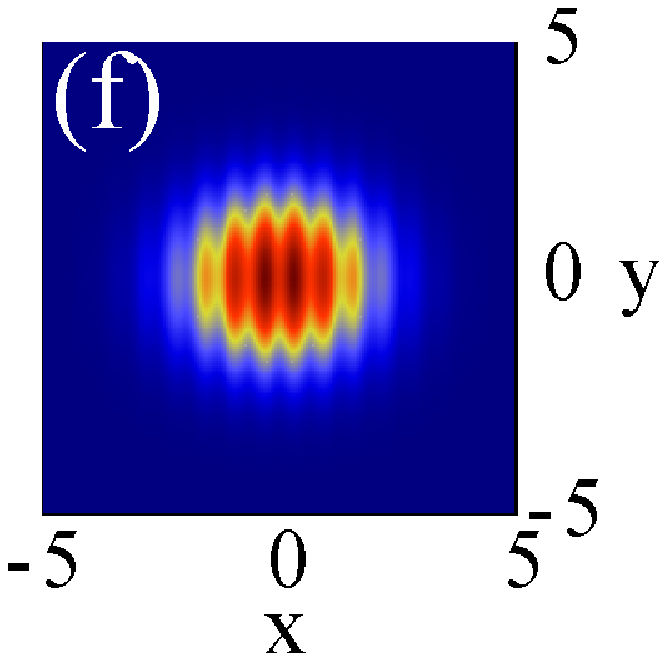}
\caption{(Color online) Density profile $|f|^{2}$ obtained from the
numerical solution of Eq. (\protect\ref{static_f}), which includes the
trapping potential (\protect\ref{POT}), for (a) $\Gamma=-1$ and $\protect%
\gamma=1$; (b) $\Gamma=-1$ and $\protect\gamma=3$; (c) $\Gamma=-1$ and $%
\protect\gamma=5$; (d) $\Gamma=1$ and $\protect\gamma=1$; (e) $\Gamma=1$ and
$\protect\gamma=3$; (f) $\Gamma=1$ and $\protect\gamma=5$. Here, we set $%
g=-1 $ and $\protect\lambda=1/3$ in all the cases. }
\label{F7}
\end{figure}

Next, we consider the same system, but in presence of the HO potential,
taken as in Eq. (\ref{POT}). Our goal is to extend the stability, with the
help of the confining potential, to $\Gamma>0$, when all the free-space
solitons are unstable. In Fig. \ref{F7} we present the numerical results
obtained for different values of the SO ($\gamma$) and Rabi ($\Gamma$)
couplings. For a fixed value of $\Gamma$ {[}e.g., $\Gamma=-1$ in Figs. \ref%
{F7}(a-c) and $\Gamma=1$ in Figs. \ref{F7}(d-f){]} and increasing $\gamma$,
we observe an increase of the number of spatial oscillations of the
solution, and, consequently, a reduction in its amplitude, as predicted by
approximated solution (\ref{b2}). Eventually, as well as in the case of the
repulsive interatomic interactions, we have numerically verified that the HO
trapping potential stabilizes the ground-state solutions displayed in Fig. %
\ref{F7}.

\section{Conclusion}

Starting from the full 3D system of the GP equations for the binary BEC,
including the SO (with equal Rashba and Dresselhaus terms) and Rabi
couplings, we have derived a system of two coupled 2D NPSEs (nonpolynomial
Schrödinger equations) for the SO-coupled BEC in the pancake-shaped
configuration. Further, assuming that the strengths of the nonlinear
interactions between different atomic states are equal, we have reduced the
stationary version of the system to the single nonlinear equation, for a
given chemical potential. This simplification has allowed us to obtain
simple approximate analytical solutions, and consider the low- and
high-density limits. By means of systematic simulations, we have obtained
localized solutions in perfect agreement with the analytical predictions. In
the case of the attractive interactions, the most essential result is
finding the \emph{stability area} for the 2D fundamental (single-peak)
solitons in the \emph{free space}, which is impossible without the SO and
Rabi couplings.

The analysis can be naturally extended by incorporating more general forms
of the SO coupling, as well as by including spatially inhomogeneous
nonlinearity, which may induce an effective nonlinear potential for solitons.

\begin{acknowledgments}
L.S. appreciates partial support from Universitá di Padova (Research Project
``Quantum Information with Ultracold Atoms in Optical Lattices" ), Cariparo
Foundation (Excellence Project ``Macroscopic Quantum Properties of Ultracold
Atoms under Optical Confinement"), and Ministero Istruzione Universita
Ricerca of Italy (PRIN Project ``Collective Quantum Phenomena: from
Strongly-Correlated Systems to Quantum Simulators"). W.B.C. thanks Brazilian
agencies CNPq and the National Institute of Science and Technology for
Quantum Information (INCT-IQ) for partial support. The work of B.A.M. was
supported, in a part, by the German-Israel Foundation through grant No.
I-1024-2.7/2009, and by the Binational Science Foundation (US-Israel)
through grant No. 2010239.
\end{acknowledgments}

\end{document}